\documentclass{emulateapj} \usepackage{natbib}

\usepackage{amsmath} 
\usepackage{graphicx}
\usepackage{color}
\graphicspath{ {./figs_psub5v0/}}
\DeclareGraphicsExtensions{.png}

\def\reff@jnl#1{{\rm#1\/}}

\def\aj{\reff@jnl{AJ}}                  
\def\araa{\reff@jnl{ARA\&A}}            
\def\apj{\reff@jnl{ApJ}}                        
\def\apjl{\reff@jnl{ApJ}}               
\def\apjs{\reff@jnl{ApJS}}              
\def\apss{\reff@jnl{Ap\&SS}}            
\def\aap{\reff@jnl{A\&A}}               
\def\aapr{\reff@jnl{A\&A~Rev.}}         
\def\aaps{\reff@jnl{A\&AS}}             
\def\baas{\reff@jnl{BAAS}}              
\def\jcap{\reff@jnl{JCAP}}              
\def\jrasc{\reff@jnl{JRASC}}            
\def\memras{\reff@jnl{MmRAS}}           
\def\mnras{\reff@jnl{MNRAS}}            
\def\physrep{\reff@jnl{Phys.Rep.}}
\def\pra{\reff@jnl{Phys.Rev.A}}         
\def\prb{\reff@jnl{Phys.Rev.B}}         
\def\prc{\reff@jnl{Phys.Rev.C}}         
\def\prd{\reff@jnl{Phys.Rev.D}}         
\def\prl{\reff@jnl{Phys.Rev.Lett}}      
\def\pasp{\reff@jnl{PASP}}              
\def\pasj{\reff@jnl{PASJ}}              
\def\skytel{\reff@jnl{S\&T}}            
\def\solphys{\reff@jnl{Solar~Phys.}}    
\def\sovast{\reff@jnl{Soviet~Ast.}}     
\def\ssr{\reff@jnl{Space~Sci.Rev.}}     
\def\nat{\reff@jnl{Nature}}             

\def\sun{\hbox{$\odot$}}

\newcommand{\hmpc}{\ensuremath{h^{-1}\mathrm{Mpc}}}

\newcommand{\hMsun}{\ensuremath{h^{-1}M_{\odot}}}

\newcommand{\beq}{\begin{equation}}
\newcommand{\eeq}{\end{equation}}
\newcommand{\beqa}{\begin{eqnarray}}
\newcommand{\eeqa}{\end{eqnarray}}

\begin{document}

\title{Impact of Baryonic Physics on Intrinsic Alignments}

\author{Ananth Tenneti\altaffilmark{1,2}, Nickolay Y.\ Gnedin\altaffilmark{1,3,4}, Yu Feng \altaffilmark{5}}

\altaffiltext{1}{Particle Astrophysics Center, Fermi National Accelerator Laboratory, Batavia, IL 60510, USA; vat@andrew.cmu.edu}
\altaffiltext{2}{McWilliams Center for Cosmology, Department of Physics, Carnegie Mellon University, Pittsburgh, PA 15213, USA}
\altaffiltext{3}{Kavli Institute for Cosmological Physics, The University of Chicago, Chicago, IL 60637 USA;}
\altaffiltext{4}{Department of Astronomy \& Astrophysics, The  University of Chicago, Chicago, IL 60637 USA} 
\altaffiltext{5}{Berkeley Center for Cosmological Physics, Department of Physics, University of California Berkeley, Berkeley, CA 94720, USA}


\begin{abstract}
We explore the effects of specific assumptions in the subgrid models of star formation and stellar and AGN feedback on intrinsic alignments of galaxies in cosmological simulations of ``MassiveBlack-II'' family. Using smaller volume simulations, we explored the parameter space of the subgrid star formation and feedback model and found remarkable robustness of the observable statistical measures to the details of subgrid physics. The one observational probe most sensitive to modeling details is the distribution of misalignment angles. We hypothesize that the amount of angular momentum carried away by the galactic wind is the primary physical quantity that controls the orientation of the stellar distribution. Our results are also consistent with a similar study by the EAGLE simulation team.
\end{abstract}


\section{Introduction}\label{S:intro}

The intrinsic shapes and orientations of galaxies are correlated with each other and the large scale density field. This intrinsic alignment of galaxies is an important astrophysical systematic in weak lensing measurements \citep{{2000MNRAS.319..649H},{2000ApJ...545..561C},{2001MNRAS.320L...7C},{2002MNRAS.335L..89J},{2004PhRvD..70f3526H}} of upcoming surveys such as the Large Synoptic Survey Telescope\footnote{\url{http://www.lsst.org/lsst/}}
(LSST; \citealt{LSST09}) and Euclid \footnote{\url{http://sci.esa.int/euclid/},
  \url{http://www.euclid-ec.org}} \citep{LAA+11}. Ignoring intrinsic alignments in weak lensing analysis can significantly bias the constraints on cosmological parameters such as the dark energy equation of state parameter \citep{2016MNRAS.456..207K}. Therefore, intrinsic alignments have been studied with analytical models and also cosmological simulations including $N$-body and hydrodynamic simulations which can help in mitigating this contaminant signal. Analytically, intrinsic alignments have been modeled with a linear alignment model \citep{{2001MNRAS.320L...7C},{2004PhRvD..70f3526H}} and modifications of the model which includes the non-linear evolution of the density field \citep{{2007NJPh....9..444B},{2015JCAP...08..015B}}. However, it is difficult to analytically describe the alignments of a galaxy's stellar component by accurately considering the physics of galaxy formation. There are also limitations to the use of $N$-body simulations as one has to populate halos with galaxies by assigning a random orientation \citep{2006MNRAS.371..750H} or employ semi-analytic methods \citep{2013MNRAS.436..819J}. Recently, intrinsic alignments of galaxies in large volume hydrodynamic simulations have been extensively studied with simulations of galaxy formation such as MassiveBlack-II \citep{2015MNRAS.450.1349K}, Horizon-AGN \citep{2014MNRAS.444.1453D}, EAGLE \citep{2015MNRAS.446..521S} and Illustris \citep{{2014Natur.509..177V},{2014MNRAS.444.1518V},{2014MNRAS.445..175G}}.  

Cosmological hydrodynamic simulations of galaxy formation are an important tool to study intrinsic alignments as it is directly possible to measure the shape and orientation of the stellar component of galaxies in the simulations. In a precursor of this paper, \cite{2015MNRAS.448.3522T} studied the galaxy shapes and two-point statistics in the MassiveBlack-II cosmological hydrodynamic simulation. This study was extended to compare the galaxy alignments based on their morphological type in MassiveBlack-II and Illustris simulations \citep{2015arXiv151007024T}. \cite{2015MNRAS.454.2736C} used the Horizon-AGN simulation, an Adaptive Mesh Refinement (AMR) based hydrodynamic simulation of galaxy formation to study intrinsic alignments of spirals and elliptical galaxies. The redshift and luminosity evolution of alignments in the same simulation was studied in \cite{2016arXiv160208373C}. Recently, \cite{2016arXiv160603216H} studied the mass and redshift dependence of intrinsic alignments in the Illustris simulation and their dependence on stellar mass, luminosity, redshift and photometric type. Qualitatively, the properties of galaxy shapes and alignments have a similar trend with mass across different simulations. However, differences have been noted in the amplitude of galaxy alignments and morphological fraction of disk galaxies in MassiveBlack-II and Illustris \citep{2015arXiv151007024T}, as well as qualitative differences in the comparison of alignments of spirals with the over-density and the redshift dependence of intrinsic alignments in the Horizon-AGN simulation \citep{{2015MNRAS.454.2736C},{2016arXiv160208373C}}. Given the differences in the models of subgrid physics adopted in these simulations and also the numerical implementations of hydrodynamics, it is important to understand the details of the subgrid physics responsible for changes in the galaxy alignments and to explore the robustness of simulation results.

In a previous study, \cite{2015MNRAS.453..721V} studied intrinsic alignments using the EAGLE suite of simulations with variations in the strength of feedback. Here, we undertake a parameter space study of the subgrid model adopted in the MassiveBlack-II simulation using a suite of small volume simulations with box size of $25h^{-1}Mpc$ on a side. We vary the free parameters in the feedback models of the simulation and test the robustness of the galaxy shapes, orientations and two-point statistics of shape correlations to variations in these parameters. Since high resolution hydrodynamic simulations of large volume are computationally expensive, we also test the usefulness of using small volume simulations to capture the sensitivity of intrinsic alignment statistics to variations in the feedback parameters. 

This paper is organized as follows. In Section~\ref{S:simulations}, we describe the simulations used in this study along with a brief overview of the feedback models adopted in the MassiveBlack-II simulation. Section~\ref{S:methods} provides the details of the methods adopted to calculate shapes and intrinsic alignment statistics studied in this paper. In Section~\ref{S:dcmode} we compare the results from the suite of small volume simulations with the fiducial MBII model and different amplitudes of the DC mode with those of the original $100h^{-1}Mpc$ box size MBII simulation. The intrinsic alignment statistics in the small volume runs with different feedback parameters are compared with those from the fiducial model in Section~\ref{S:baryonic}. Finally, we provide a summary of our conclusions in Section~\ref{S:conclusions} 

\section {Simulations and Feedback Models} \label{S:simulations}

In this paper, we use the MassiveBlack-II (MBII) simulation \citep{2015MNRAS.450.1349K}, a high resolution cosmological hydrodynamic simulation performed in a box of volume $(100h^{-1}Mpc)^3$, which includes galaxy formation physics as our base model. We complement MassiveBlack-II with smaller volume simulations of size $25h^{-1}Mpc$, in which we vary the key parameters of the star formation and stellar and AGN  feedback model. We denote the smaller volume simulations as MBII-25. The simulations are performed with the TreePM-Smoothed Particle Hydrodynamics (SPH) code, P-Gadget, a modified version of GADGET2 \citep{2005MNRAS.361..776S}. The same version of the code has been used earlier to perform the large volume MBII simulation \citep{2015MNRAS.450.1349K}. The simulations include the wide range of physical effects thought to be crucial for properly modeling galaxy formation, such as multiphase ISM, star formation, supernova and stellar wind feedback, as well as black hole accretion and feedback. Radiative cooling and heating are included as in \cite{1996ApJS..105...19K}, along with photoheating due to an imposed ionizing UV background. 

Initial conditions are generated at $z=159$ and simulations are evolved to $z=0$ with an equal initial number of gas and dark matter particles. The cosmological parameters are chosen with the WMAP7 cosmology\citep{2011ApJS..192...18K}: $h=0.701$, $\Omega_{m}=0.275$, $\Omega _{b} = 0.046$, $\Omega _{\Lambda} = 0.725$, $\sigma _{8} =0.816$, spectral index, $\eta _{s} = 0.968$  The mass of each dark matter particle is $1.1\times 10^{7}h^{-1}M_{\odot}$. The smaller volume simulations are performed with the same mass and spatial resolution as the original simulation. Accordingly, the initial number of gas and dark matter particles are equal to $2 \times 1792^{3}$ and $2 \times 448^{3}$ in the $100h^{-1}Mpc$ and $25h^{-1}Mpc$ box size simulations respectively. We note that all the small volume simulations have been started with the same initial conditions at $z=159$. The details of the star formation and feedback models of the simulation and the changes adopted in the small volume runs are described below. 

\subsection{Star formation and Stellar and AGN Feedback } \label{SS:sf}

The star formation and feedback model adopted in the simulation is based on an earlier multiphase ISM model of Springel \& Hernquist \cite{2003MNRAS.339..289S}. Specifically, if the local gas density $\rho$ is greater than a critical density threshold $\rho _{th}$, a multiphase ISM consisting of cold clouds in pressure equilibrium with a hot ambient gas is assumed. The effective pressure $P_{eff}$ is defined as $P_{eff} = (\gamma - 1)(\rho _{h} \mu _{h} + \rho _{c} \mu _{c})$ \citep{2003MNRAS.339..289S}, where $\rho _{c}$, $\rho _{h}$ are the local densities of cold and hot phases respectively, $\rho = \rho _{c} + \rho _{h}$, and $\mu _{h}$ and $\mu _{c}$ are specific energies of hot and cold components. The threshold density $\rho _{th}$ is determined self consistently by requiring that the effective pressure is a continuous function of density.

Star formation is modeled by spawning individual stellar particles stochastically from the cold clouds. The rate of star formation is given by
\begin{equation} \label{eq:starfor}
\frac{d\rho _{*}}{dt} = \frac{\rho _{c}}{t_{*}} - \beta \frac{\rho _{c}}{t_{*}}
\end{equation} 
where $\beta = 0.1$ is the mass fraction of short lived stars and $t_{*}$ is the star formation time scale with density dependence given by
\begin{equation}
t_{*}(\rho) = t_{0}^{*}(\frac{\rho}{\rho_{th}})^{-0.5},
\end{equation} 
where $t_{0}^{*} = 2.1\mathrm{Gyr}$.

The energy released by supernovae heats the ambient gas and the heating rate is set by the energy balance condition
\begin{equation} \label{eq:sf_hrate}
\frac{d}{dt}(\rho _{h} \mu _{h}) = \beta \frac{\rho _{c}}{t_{*}}(\mu _{SN}).
\end{equation}
Here $\mu _{SN}=\frac{3}{2}kT_{SN}$ where $T_{SN}$ is the equivalent supernova temperature which is equal to $10^{8}$ K in the fiducial model. 

\subsection{Wind Feedback} \label{SS:wf}

Galactic winds are implemented with the wind velocity given by
\begin{equation}
v_{w} = \sqrt{\frac{2\beta \chi \mu _{SN}}{\eta (1-\beta)}},
\end{equation}
where $\chi = 1.0$ is the fraction of supernova energy carried by the wind and $\eta = 2.0$ is the wind loading factor. For a given time step $\Delta t$, a gas particle is added to the wind probabilistically with the probability
\begin{equation}
p_{w} = 1 - \exp{[-\frac{\eta (1-\beta)x\Delta t}{t_{*}}]}.
\end{equation}


\subsection{AGN Feedback} \label{SS:agn}

The simulations also include the physics of black hole accretion and feedback, based on the models of \cite{2005MNRAS.361..776S} and \cite{2005Natur.433..604D}. Black holes are treated as collisionless particles introduced into halos of mass greater than $5.0\times10^{10}h^{-1}M_{\odot}$ at regular time intervals, separated by $\Delta \log (a) = \log (1.25)$. The densest particle is converted into a seed black hole of mass $M_{\mathrm{BH,seed}} = 5\times10^{5}h^{-1}M_{\odot}$ which grows in mass by black hole accretion and mergers. The black hole accretion rate is given by the modified Bondi rate formula
\begin{equation}
\dot{M}_{BH} = \frac{4\pi \alpha G^{2}M^{2}_{\mathrm{BH}}\rho}{(c^{2}_{s} + v^{2}_{\mathrm{BH}})^{3/2}},
\end{equation}
where $\rho$ is the local gas density, $c_{s}$ is the local speed of sound, $v$ is the velocity of BH relative to the gas. The accretion rate is limited to $2$ times the Eddington rate, $\dot{M}_{\mathrm{Edd}}$. A dimensionless parameter $\alpha$ is set to 100; that value has been found experimentally to approximately correct for the gas density close to the black hole, which is reduced in the effective sub-resolution model of the ISM.

The AGN feedback is modeled by coupling $5\%$ (the value chosen to match the slope in the observed $M_{\mathrm{BH}}-\sigma$ relation \citep{2005MNRAS.361..776S}) of the bolometric luminosity radiated from the BH,
\begin{equation} 
L_{\mathrm{bol}}= \epsilon _{r}\dot{M}_{\mathrm{BH}}c^{2},
\end{equation}
with the radiation efficiency $\epsilon _{r}=0.1$. The energy is deposited isotropically to the 64 nearest gas particles within the BH particle kernel.  

\subsection{Parameters Space Study} \label{SS:feedbackpar}

In the simulations analyzed here, we vary the key parameters in the star formation and stellar and AGN feedback models. In particular, we consider the effect of a lower or higher star formation efficiency by increasing and decreasing the star formation timescale $t_{0}^{*}$ by a factor of 3. We also consider the effects of increasing the AGN feedback by increasing the scaling parameter $\alpha$ in the AGN feedback model to $300$, which triples the black hole accretion rate. Similarly, the effect of wind velocity is weakened by decreasing the wind loading factor 10 times to study the effects of wind feedback. 

\section{Methods} \label{S:methods}

In this section, we describe the method adopted to calculate shapes and the also provide details of the intrinsic alignment statistics explored in this paper. 

\subsection{Calculation of shapes}

The 3D shapes of the dark matter and stellar components in subhalos are determined using the
the eigenvalues and eigenvectors of the reduced inertia tensor given by
\begin{equation} \label{eq:redinertensor}
\widetilde{I}_{ij} = \frac{\sum_{n} m_{n}\left(x_{ni}x_{nj}\right)/r_{n}^{2}}{\sum_{n} m_{n}},
\end{equation}
where the summation is over particles index $n$, and 
\begin{equation} \label{eq:rn2}
 r_{n}^{2} = \frac{x_{n0}^{2}}{a^{2}} + \frac{x_{n1}^{2}}{b^{2}} + \frac{x_{n2}^{2}}{c^{2}}.
\end{equation}
Here $a$, $b$, and $c$ are half-lengths of the principal axes of the ellipsoid. 

The eigenvectors of the inertia tensor are
${\hat{e}_{a}, \hat{e}_{b}, \hat{e}_{c}}$ with corresponding
eigenvalues 
$\lambda_{a} > \lambda_{b} > \lambda_{c}$. The eigenvectors represent
the principal axes of the ellipsoid, with the half-lengths of the principal
axes $(a,b,c)$ given by 
$(\sqrt{\lambda_{a}},\sqrt{\lambda_{b}},\sqrt{\lambda_{c}})$. The 3D
axis ratios are $b/a$ and $c/a$.

Similarly, in 2D, the
projected shapes are calculated by projecting the positions of the particles
onto the $XY$ plane and modeling the shapes as ellipses. Here, we denote the eigenvectors as ${\hat{e}_{a}',\hat{e}_{b}'}$ with 
corresponding eigenvalues 
${\lambda_{a}' > \lambda_{b}'}$. The lengths of the semi-major and semi-minor axes
are $a' = \sqrt{\lambda_{a}'}$ and $b' = \sqrt{\lambda_{b}'}$ with the axis
ratio $b'/a'$.

The details of the iterative method for measuring axis ratios can be found in \cite{2015MNRAS.448.3522T}. In the first iteration, we start with the half-lengths of the principal axes all equal to $1$ and determine the eigenvalues and eigenvectors of the ellipsoid. After each iteration, the lengths
of the principal axes of ellipsoids are rescaled such that the enclosed volume is constant and particles outside the ellipsoidal volume are discarded. This process is repeated until convergence is
reached such that the fractional change in
axis ratios is below 1\%. 

In addition to the distribution of the axis ratios, $b/a$ and $c/a$ of the stellar components of subhalos, we are also interested in the orientation of the major axis of the stellar shape with the shape of dark matter in subhalos. So, we compute 
the probability distribution of the misalignment angle
\begin{equation} \label{eq:misalignangle}
 \theta_{m} = \arccos(\left|\hat{e}_{da} \cdot \hat{e}_{ga}\right|),
\end{equation}
where $\hat{e}_{da}$ and $\hat{e}_{ga}$ are the major axes of the shapes
defined by the dark matter and stellar matter components respectively.

\subsection{Two-point statistics}

In this paper we quantify the intrinsic alignments of galaxies with the large-scale
density field using the ellipticity-direction (ED) and the
projected shape-density ($w_{\delta +}$) correlation functions. 

The ED correlation function cross-correlates the orientation of the
major axes of the 3D shapes of dark matter or stellar component of galaxies with the large-scale density
field. Consider a subhalo centered at position \textbf{x} with the major axis
direction $\hat{e}_{a}$. Let the unit vector in the direction of a tracer of the matter density field at a distance $r$ be $\hat{\textbf{r}}(\textbf{x})$. Based on the notation in \cite{2008MNRAS.389.1266L},
the ED correlation function is given by
\begin{equation} \label{eq:ED3d}
 \omega _{\delta} (r) = \langle \mid \hat{e}_{a}(\textbf{x})\cdot \hat{\textbf{r}}(\textbf{x}) \mid^{2} \rangle - \frac{1}{3},
\end{equation}
which is zero for randomly oriented galaxies in a uniform distribution. In the simulations the matter density field is traced using the positions of dark matter particles.

The projected shape correlation function, $w_{\delta +}$ is directly related to the correlation function measured in observations. Following the notation of 
\cite{2006MNRAS.367..611M}, we define the
the matter-intrinsic shear correlation function
$\hat{\xi}_{\delta +}(r_{p},\Pi)$ and the corresponding projected
two-point statistic $w_{\delta +}$. In this paper, $r_{p}$ is the comoving
transverse separation of a pair of galaxies in the $XY$ plane and
$\Pi$ is their separation along the $Z$ direction.

The components of the projected ellipticities of a galaxy are given by 
\begin{equation} \label{eq:ellipticity}
 (e_{+},e_{\times}) = \frac{1 - (b'/a')^{2}}{1 +
    (b'/a')^{2}}\left[\cos{(2\phi)},\sin{(2\phi)}\right],
\end{equation}
where $b'/a'$ is the axis ratio of the projected shape of the 
stellar component of a galaxy, and $\phi$ is the position angle 
of the major axis with respect to the reference direction (position of
the dark matter particle). Here, $e_{+}$ refers to the radial component and
$e_{\times}$ is the component rotated at $45^{\circ}$.  The
matter-intrinsic shear correlation function is given by,
\begin{equation} \label{eq:gicorr}
 \hat{\xi}_{\delta +}(r_{p},\Pi) = \frac{S_{+}D}{RR}
\end{equation}
where $S_{+}$ represents the ``shape sample'', selected on the basis of a binning in subhalo mass and the ``density sample'' labeled by $D$ consists of the dark matter particles used to trace the matter density field. $S_{+}D$ is given by the following sum over all galaxy - dark matter particle pairs with separations $r_{p}$ and $\Pi$:
\begin{equation} \label{eq:SpD}
 S_{+}D = \sum_{i\neq j\mid r_{p},\Pi}\frac{e_{+}(j\mid
   i)}{2\mathcal{R}},
\end{equation}
where $e_{+}(j | i)$ is the $+$ component of the ellipticity of a
galaxy ($j$) from the shape sample relative to the direction of a dark 
matter particle ($i$) selected from the density sample. Here, $\mathcal{R}
= (1 - e_\text{rms}^{2})$ is the shear responsivity that converts from distortion
to shear \citep{2002AJ....123..583B}, with $e_\text{rms}$ being the RMS ellipticity per component of the shape sample. The $RR$ term in Eq.~\eqref{eq:gicorr} refers to the
expected number of randomly-distributed pairs in a particular $(r_p, \Pi)$ bin around galaxies in the shape sample. 

The projected shape correlation function 
$w_{\delta +}(r_{p})$ is 
given by
\begin{equation} \label{eq:wgp}
 w_{\delta +}(r_{p}) =
 \int_{-\Pi_\text{max}}^{+\Pi_\text{max}}\hat{\xi}_{\delta +}(r_{p},\Pi)\,\mathrm{d}\Pi.
\end{equation}
We calculate the matter-intrinsic shear correlation function over the whole length
of the box, $L_{box}$ with $\Pi_\text{max} = L_{box}/2$, where the length of
  the box is $100\hmpc$ or $25 \hmpc$. The projected
correlation functions are obtained via direct summation. 

\section{Intrinsic alignments in a smaller volume box including DC mode in the fiducial model} \label{S:dcmode}

To study the effects of modifying baryonic feedback parameters, we use small volume simulations, as larger simulation volumes would not be feasible at present. Smaller volume simulations, however, will be a subject to larger cosmic variance, and so may be biased relative to the larger box.

In order to estimate the error we are going to incur by using smaller boxes, we use the DC mode formalism \citep{2005ApJ...634..728S} that allows one to approximately quantify the effect of the missing large-scale power. Ideally, one would need to run a whole ensemble of the simulations with randomly chosen DC modes. However, due to limited computational resources, we only perform three independent realizations of the $25h^{-1}Mpc$ box with the amplitude of DC mode set to zero and to $\pm \Delta_{0}$, where $\Delta_{0}$ is the rms density fluctuations in a cubic $25h^{-1}Mpc$ box at $z=0$. For any of our statistical measures we then can use the spread between the three realizations as an, admittedly crude, estimate of the uncertainty due to the limited simulation volume.

For the WMAP7 cosmological parameters and the box size of $25h^{-1}Mpc$ box size $\Delta _{0} = 0.585$. In a most general case accounting for the DC mode requires modifications to the simulation code. However, \citep{2005ApJ...634..728S} showed that for the cosmology that includes only matter, the cosmological constant, and, optionally, curvature, the DC mode can be accounted for by a simple rescaling of cosmological parameters. In this paper we use such a rescaling to include the DC mode in P-Gadget that does not support the DC mode explicitly.

\subsection{Distribution of Shapes and Misalignment angles}

\begin{figure}[t]
\includegraphics[width=0.5\hsize]{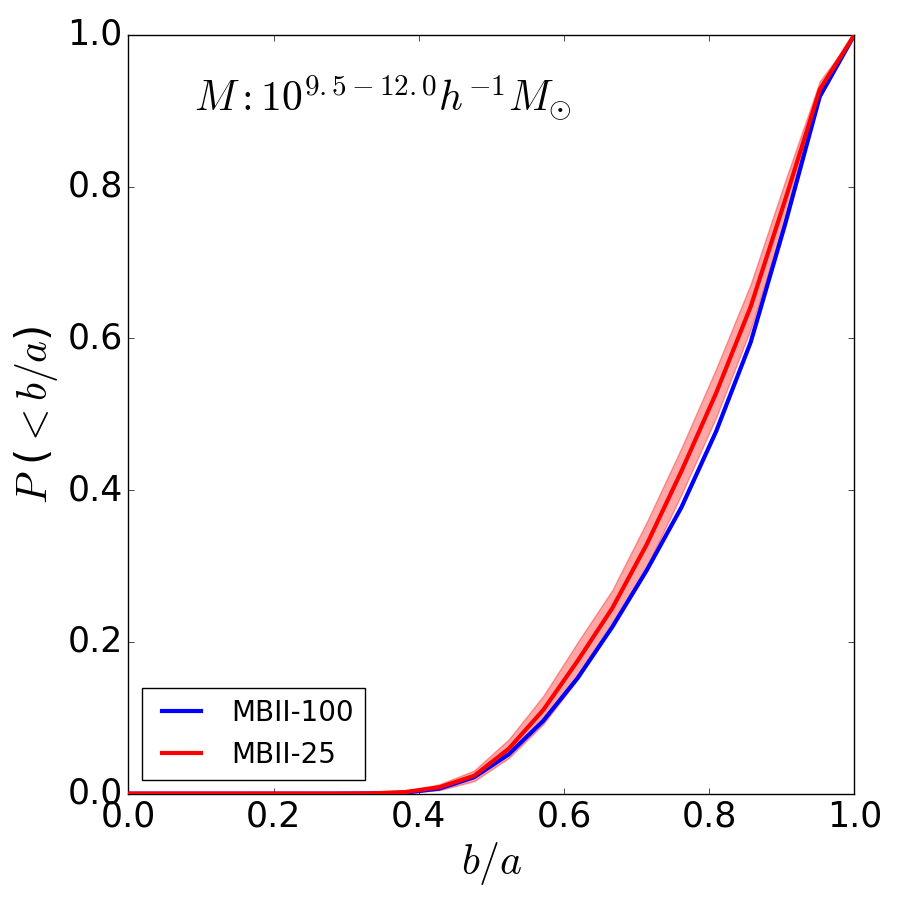}%
\includegraphics[width=0.5\hsize]{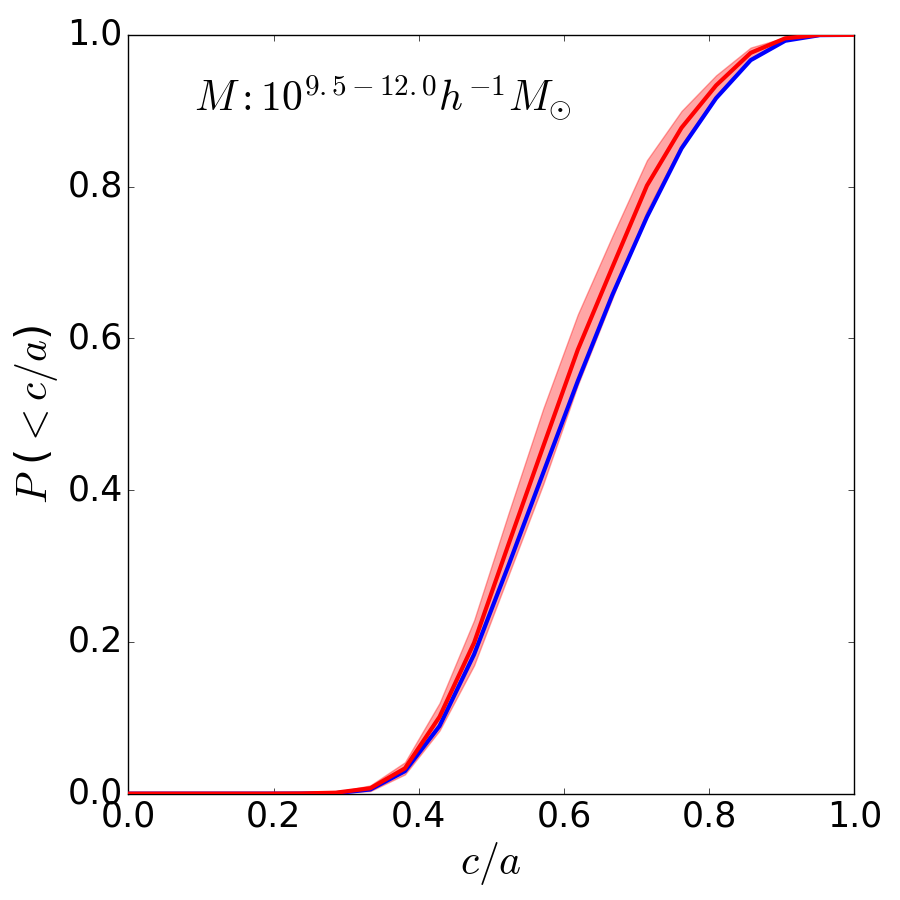}\newline%
\includegraphics[width=0.5\hsize]{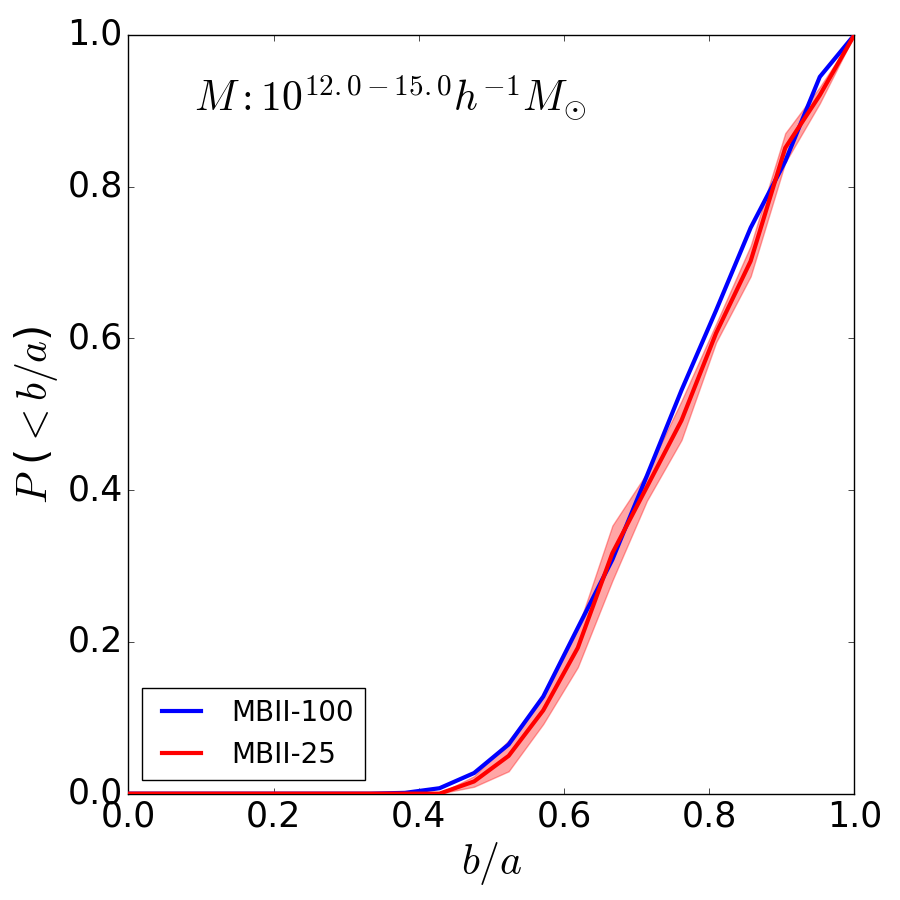}%
\includegraphics[width=0.5\hsize]{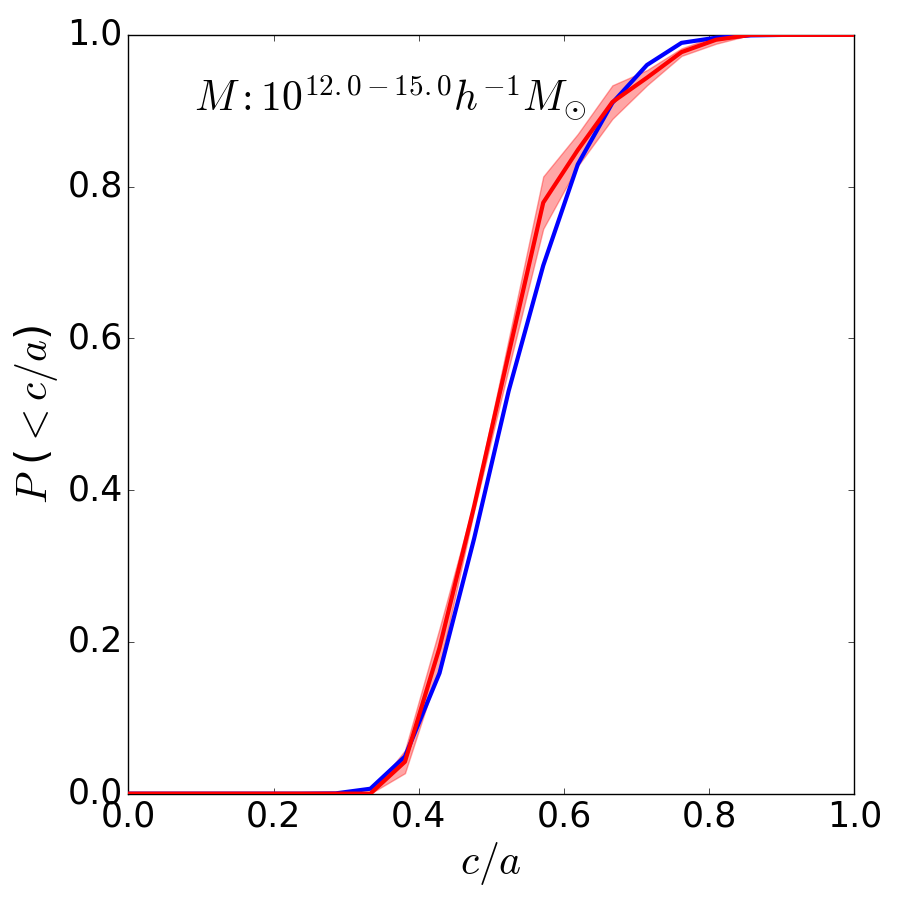}%
\caption{\label{F:fig1dcm_qs} Cumulative distribution function (CDF) of the shapes $b/a$ and $c/a$ in two mass bins $10^{9.5-12.0}h^{-1}M_{\odot}$ and $10^{12.0-15.0}h^{-1}M_{\odot}$ of MBII-100 run and the mean CDF of the shapes of three MBII-25 simulations with different DC modes. The bands show the error of the mean CDF.}
\end{figure}

\begin{table*}
\begin{center}
\caption{\label{T:tab1} Mean 3D shapes $b/a$ and $c/a$ of the stellar component of the fiducial MBII-100 simulation and three MBII-25 simulations with different DC modes of 0 and $\pm1\sigma$}
\begin{tabular}{@{}lcccccccc}
\hline
 & \multicolumn{2}{c}{MBII-100} & \multicolumn{2}{c}{MBII-25, $\Delta_{\rm DC}=0$} & \multicolumn{2}{c}{MBII-25, $\Delta_{\rm DC}=+1\sigma$} & \multicolumn{2}{c}{MBII-25, $\Delta_{\rm DC}=-1\sigma$}\\
\hline
 $M_{subhalo}$ (\hMsun) & $\langle b/a \rangle$ & $\langle c/a \rangle$ & $\langle b/a \rangle$ & $\langle c/a \rangle$ & $\langle b/a \rangle$ & $\langle c/a \rangle$ & $\langle b/a \rangle$ & $\langle c/a \rangle$\\
\hline
 $10^{9.5}-10^{12.0}$ & $0.79 \pm 0.0$ & $0.61 \pm 0.0$ & $0.78 \pm 0.0$ & $0.60 \pm 0.0$ & $0.75 \pm 0.0$ & $0.56 \pm 0.0$ & $0.795 \pm 0.003$ & $0.62 \pm 0.0$\\
 $10^{12.0}-10^{15.0}$ & $0.74 \pm 0.0$ & $0.525 \pm 0.002$ & $0.76 \pm 0.02$ & $0.524 \pm 0.015$ & $0.73 \pm 0.02$ & $0.515 \pm 0.013$ & $0.77 \pm 0.02$ & $0.51 \pm 0.015$\\
\end{tabular}
\end{center}
\end{table*}

In Figure \ref{F:fig1dcm_qs} we show a comparison between the
cumulative distribution functions (CDF) for the shapes, $b/a$ and
$c/a$ in two mass bin for the original $100h^{-1}Mpc$ MBII-100 run and
our three $25h^{-1}Mpc$ MBII-25 simulations with different DC
modes. The mean values for the shapes are tabulated in Table
\ref{T:tab1}. Because of the limited size of our simulation volumes,
we are only able to consider two mass bins. However, this may be
sufficient to notice a really strong trend with halo mass; more subtle
trends are missed by us and will have to be explored in the future
with more precise simulations. Throughout this paper, the galaxy shapes
and alignments are analyzed at $z=0.3$. 

\begin{figure}[t]
\begin{center}
\includegraphics[width=0.5\hsize]{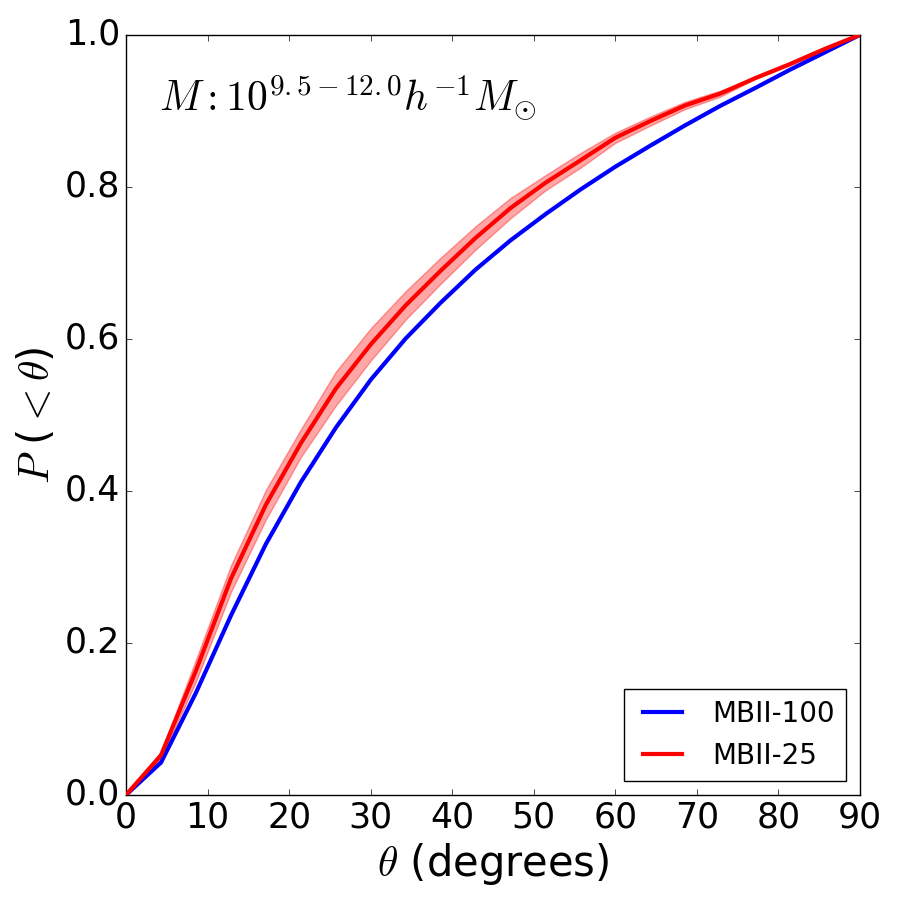}%
\includegraphics[width=0.5\hsize]{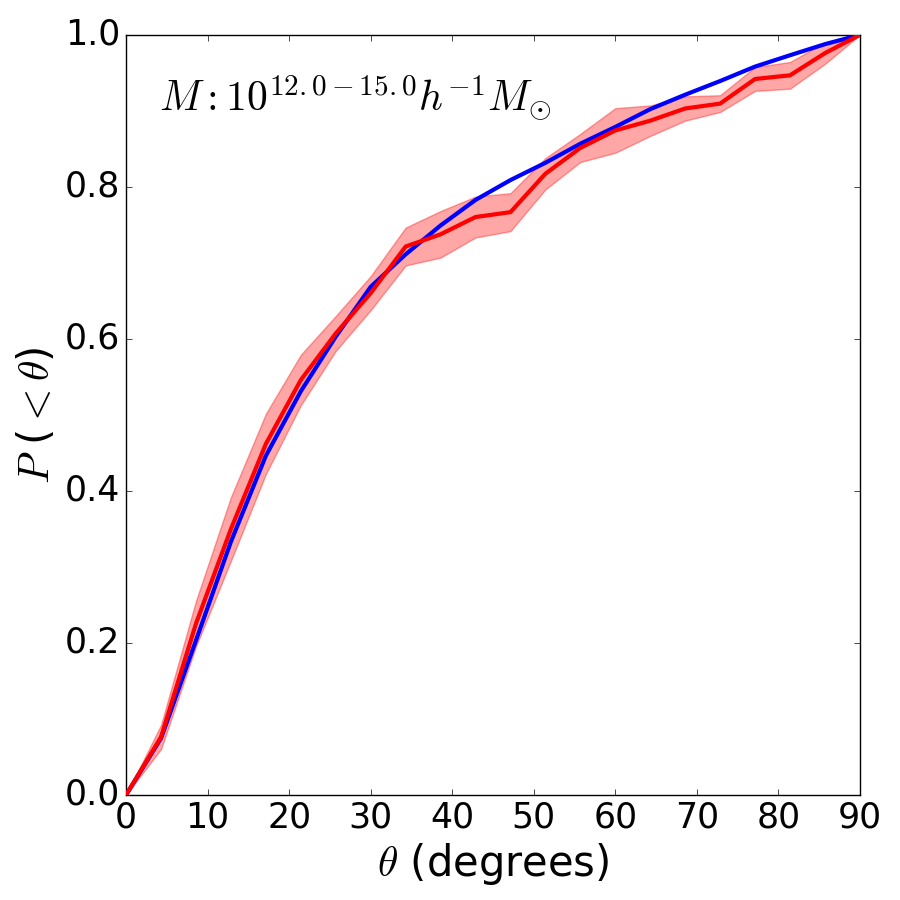}%
\caption{\label{F:fig3dcm_ma} Cumulative distribution function (CDF) of the misalignment angle $\theta$ in the mass bins $10^{9.5-12.0}h^{-1}M_{\odot}$  and $10^{12.0-15.0}h^{-1}M_{\odot}$ of MBII $100h^{-1}Mpc$ and the mean CDF of the misalignment angles of $25h^{-1}Mpc$ simulations with different DC modes. The error on the mean CDF is indicated by the bands.
}
\end{center}
\end{figure}

\begin{table*}
\begin{center}
\caption{\label{T:tab2} Mean 3D misalignment angles, $\langle \theta \rangle$ (degrees), between the
  major axis of galaxies and their host dark matter subhalos in the  MBII simulation of $100h^{-1}Mpc$ size box and simulations of $25h^{-1}Mpc$ box with DC-modes : 0, $\pm1\sigma$}
\begin{tabular}{@{}lcccc}
\hline
 $M_{subhalo}$ (\hMsun) & MBII-100 & MBII-25, $\Delta_{\rm DC}=0$  & MBII-25, $\Delta_{\rm DC}=+1\sigma$ & MBII-25, $\Delta_{\rm DC}=-1\sigma$\\
\hline
$10^{9.5}-10^{12.0}$ & $33.259 \pm 0.078^{\circ}$ & $31.151 \pm 0.603^{\circ}$ & $31.761 \pm 0.662^{\circ}$ & $28.153 \pm 0.545^{\circ}$ \\
$10^{12.0}-10^{15.0}$ & $27.157 \pm 0.409^{\circ}$ & $31.785 \pm 3.525^{\circ}$ & $25.505 \pm 2.701^{\circ}$ & $26.49 \pm 4.35^{\circ}$ \\
\end{tabular}
\end{center}
\end{table*}

For three smaller volume simulations we can both the compute the mean over the three realization, and the error in that mean, which we show in these and all subsequent figures with lines and bands respectively. Since the small box simulations may be biased and/or insufficiently accurate, we use the error in the mean as the estimate of our theoretical error due to the limited box size. For example, from Fig.\ \ref{F:fig1dcm_qs} it is clear that the differences between the mean of three MBII-25 runs and the original MBII-100 run are comparable to the error on MBII-25, and that error is reasonably modest, about 2\%. Hence, by using smaller boxes we do introduce a bias, but the bias is modest and is comparable to the statistical error of the simulation results.

The distributions of misalignment angles in the same two mass bins are shown in Figure \ref{F:fig3dcm_ma}, and their mean values are given in Table \ref{T:tab2}. We find that the galaxies in the lower mass bin of smaller volume simulations are more aligned, at about $2\sigma$ level, than in the fiducial MBII-100 run, and in the high mass bins low abundance of halos becomes appreciable. In both cases, however, the bias in using smaller boxes is still sufficiently modest (less than $3^o$) to justify our use of smaller boxes in this first, exploratory work.

\begin{figure}
\begin{center}
\includegraphics[width=0.5\hsize]{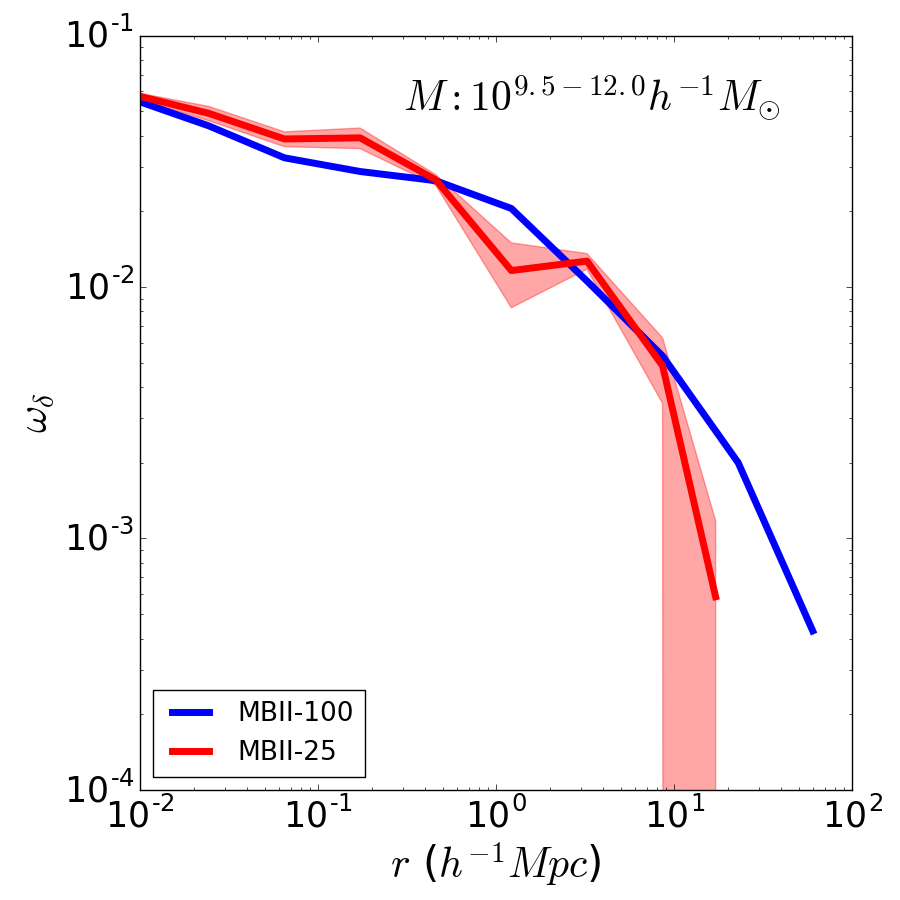}%
\includegraphics[width=0.5\hsize]{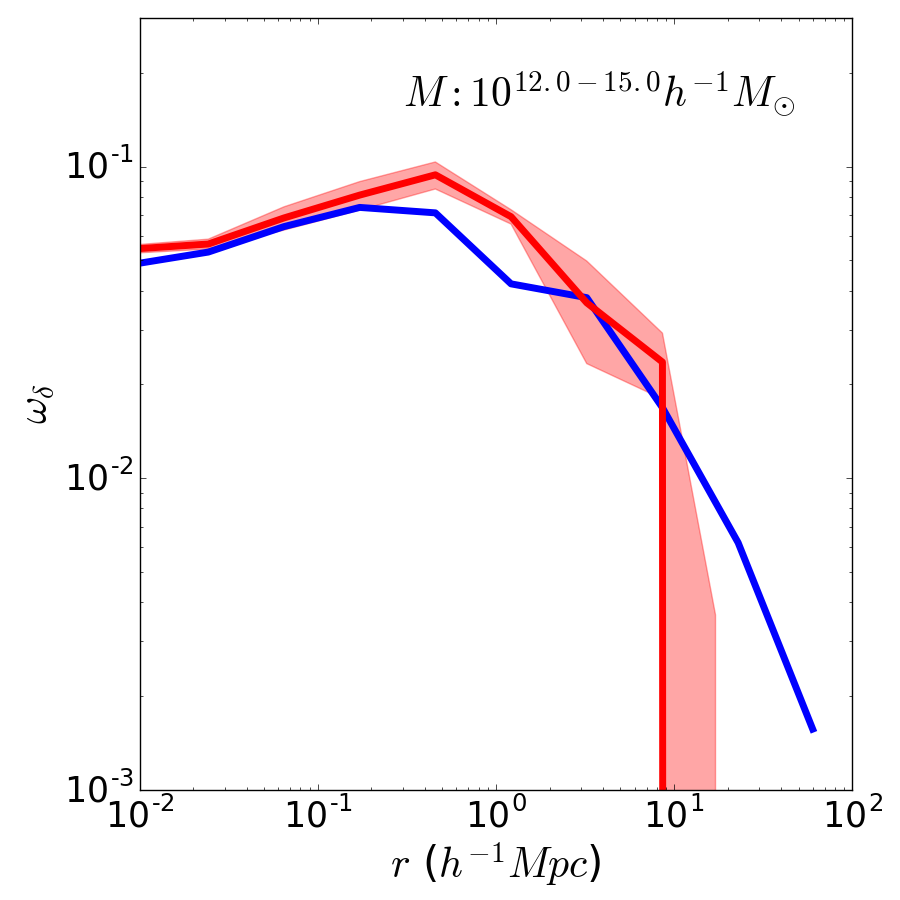}\newline%
\includegraphics[width=0.5\hsize]{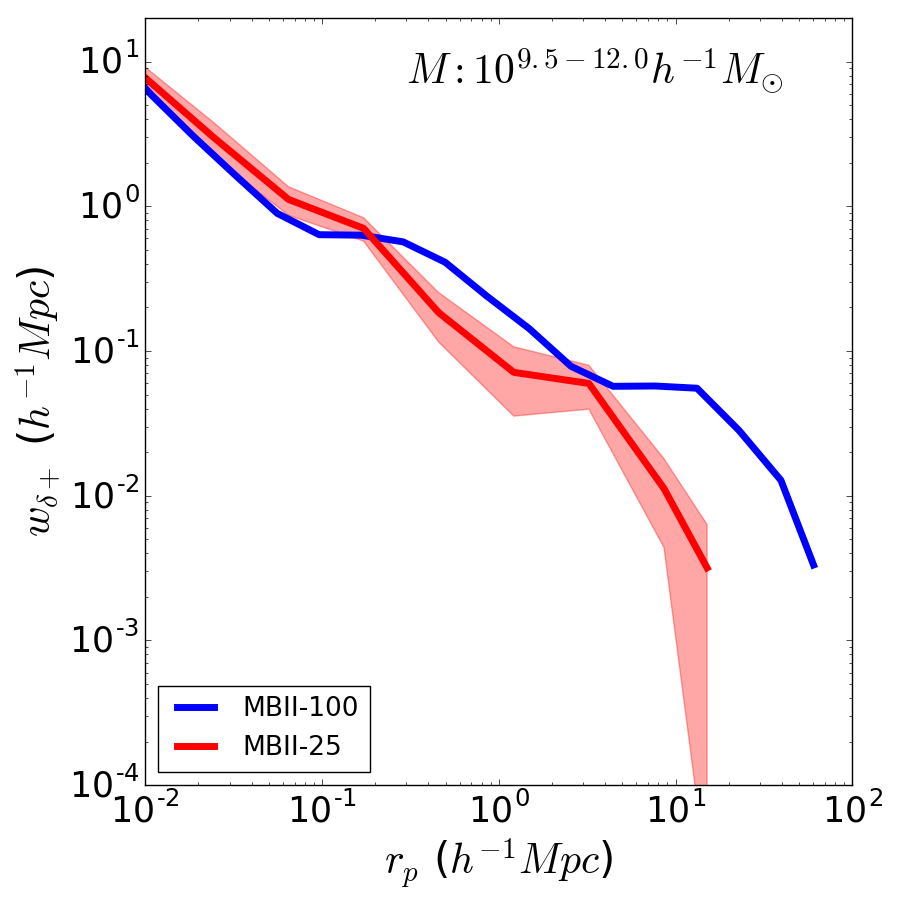}%
\includegraphics[width=0.5\hsize]{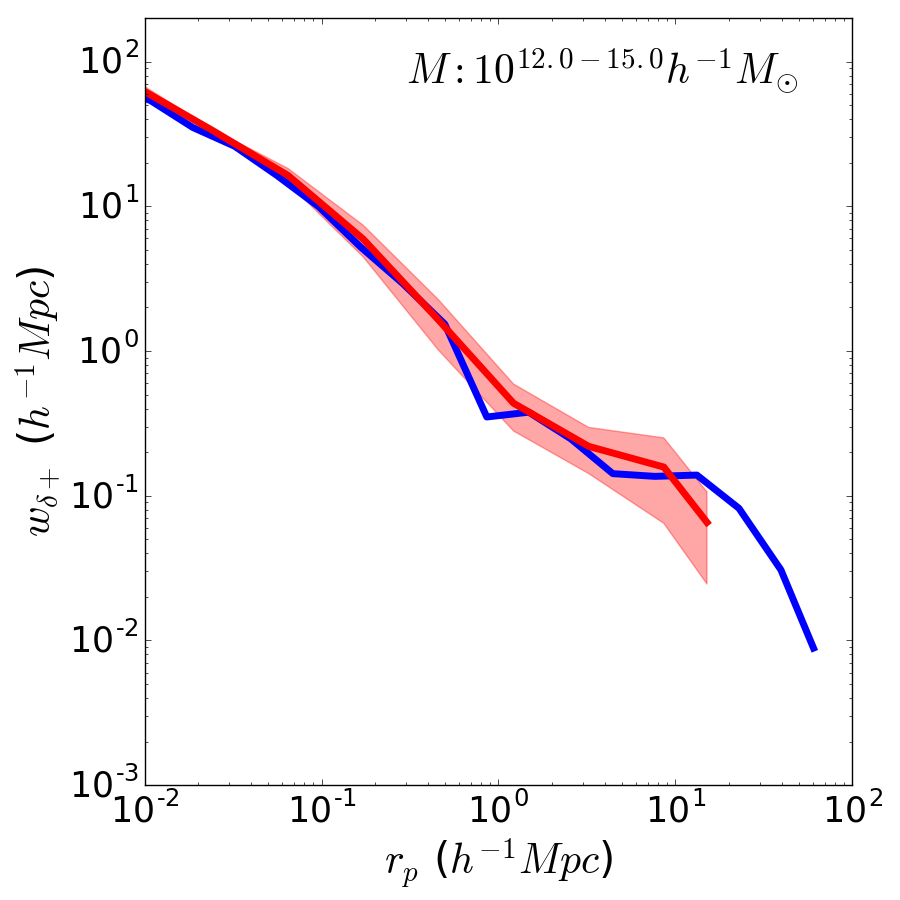}%
\caption{\label{F:fig4eddcm} ED and $w_{\delta +}$ correlation functions in two mass bins, $10^{9.5-12.0}h^{-1}M_{\odot}$ and $10^{12.0-15.0}h^{-1}M_{\odot}$, of MBII-100 and of three independent realizations of MBII-25 box with different DC modes. The bands indicate the error in the mean ED correlation function.
}
\end{center}
\end{figure}

The two-point statistics ED and $w_{\delta +}$ are shown in Figure \ref{F:fig4eddcm}. The ED and $w_{\delta +}$ correlation functions in the MBII-100 simulation and in the mean of MBII-25 runs are in good agreement on small scales  and in the high mass bin. The agreement is worse at large scales in the low mass bin, but the measurements there are also noisy. The formal error on the mean of three MBII-25 runs is smaller than the difference between the two box sizes, but since the error is estimated from just three runs, it may itself be inaccurate. 

Overall, we find that our $25h^{-1}Mpc$ boxes are a suitable, albeit not ideal and moderately biased, tool for exploring the sensitivity of the simulation predictions to the parameters of the star formation and feedback model.

\section{Baryonic effects : parameter variation in the fiducial model} \label{S:baryonic}

In this section, we explore the effects of modifying the feedback parameters in the simulation on the galaxy shapes and two-point statistics. We follow the methodology of the previous section, and use the three MBII-25 runs with different DC models as our new fiducial simulation set against which we compare runs with varied physics. The details about which parameter is varied in a given model are provided in Section~\ref{SS:feedbackpar}.

\begin{figure}
\begin{center}
\includegraphics[width=0.5\hsize]{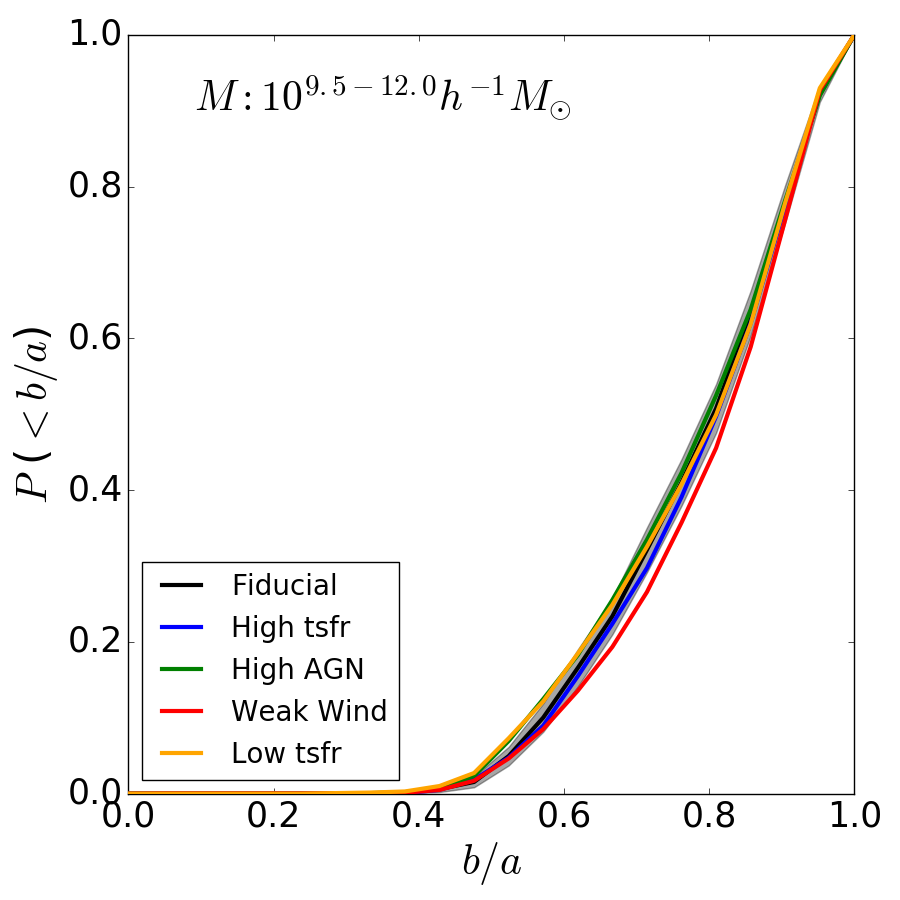}%
\includegraphics[width=0.5\hsize]{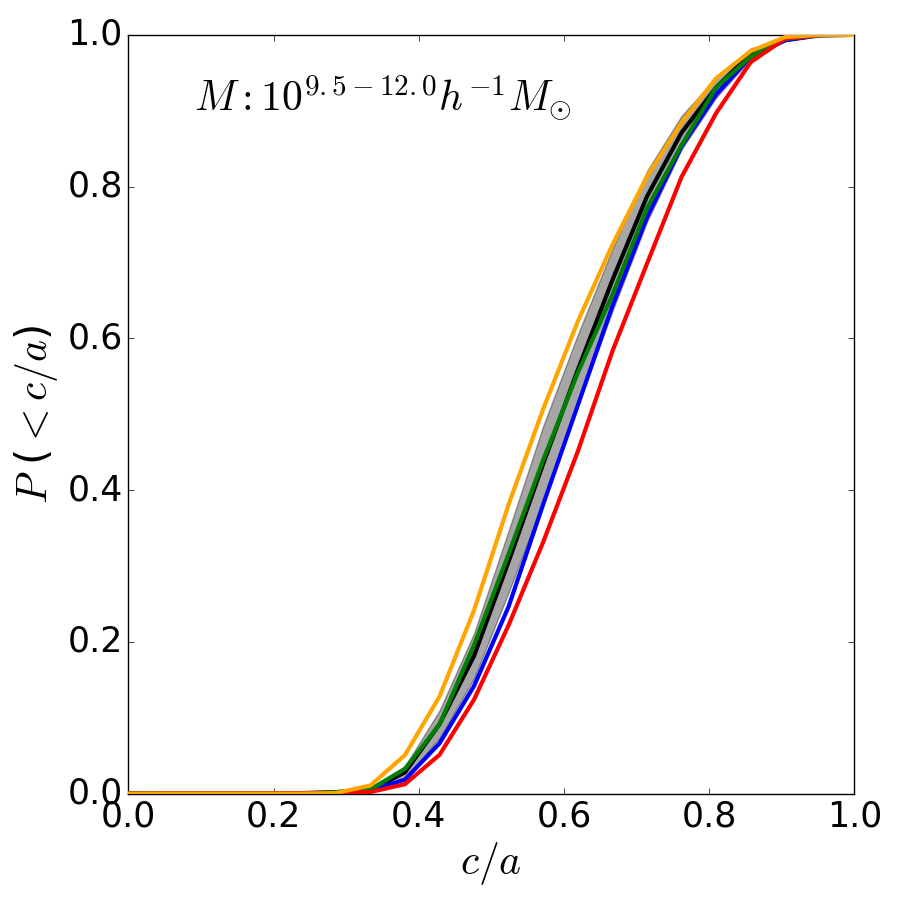}\newline%
\includegraphics[width=0.5\hsize]{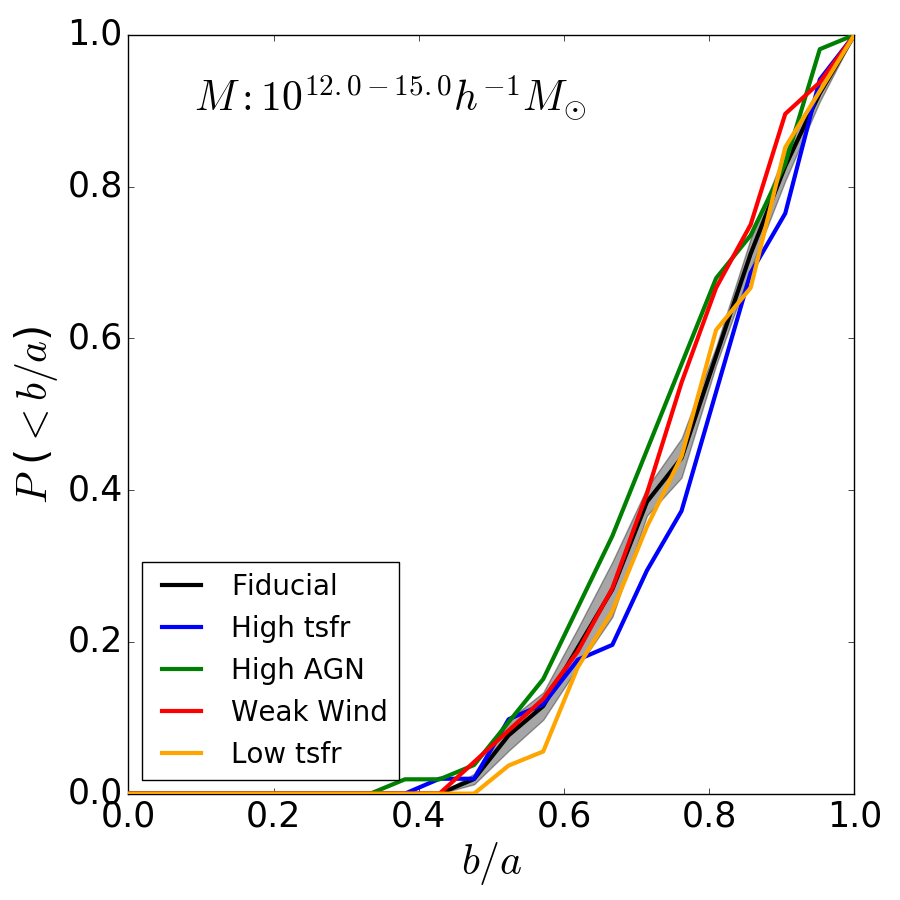}%
\includegraphics[width=0.5\hsize]{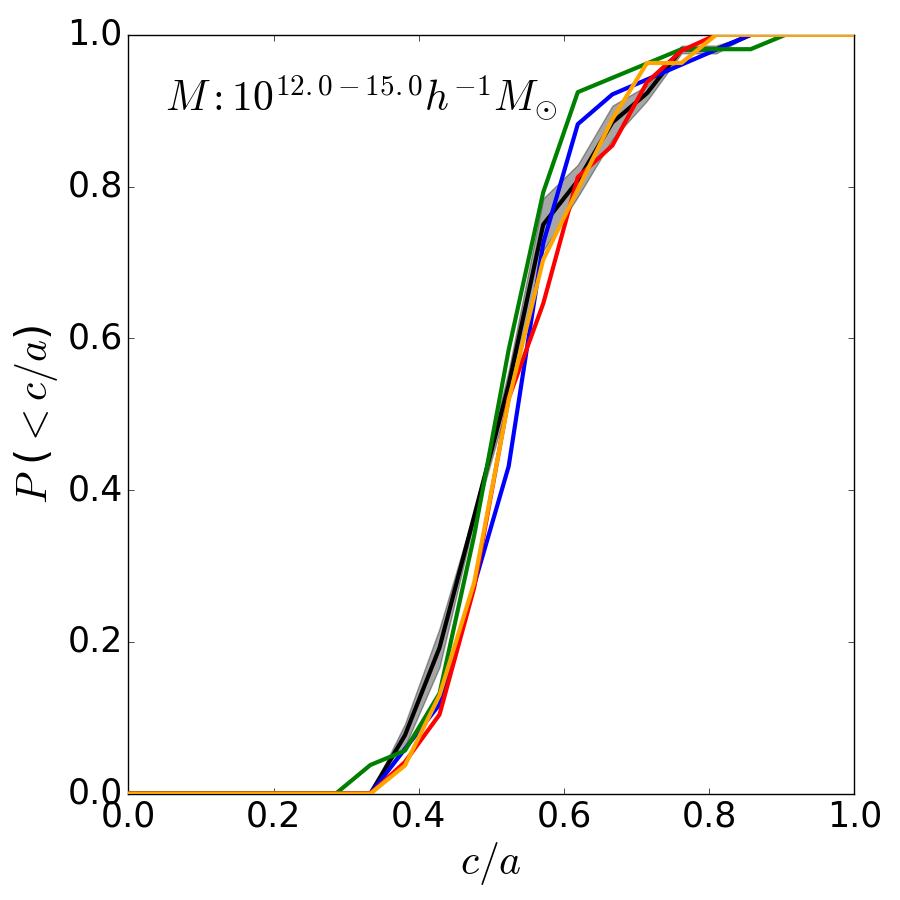}%
\caption{\label{F:fig1_qs} Cumulative distribution functions of the shapes $b/a$ and $c/a$ in two mass bins, $10^{9.5-12.0}h^{-1}M_{\odot}$ and $10^{12.0-15.0}h^{-1}M_{\odot}$, of several $25h^{-1}Mpc$ box simulations with varied physics. Black line with the gray band is the fiducial MBII-25 model and its error, shown with red lines in the previous section.
}
\end{center}
\end{figure}

\begin{table*}
\begin{center}
\caption{\label{T:tab4}  Mean of $b/a$, $c/a$, and $\theta$ of the stellar shape of galaxies for simulations with varying star formation feedback.}
\begin{tabular}{@{}lcccccc}
\hline
 & \multicolumn{3}{c}{$10^{9.5}-10^{12.0}$(\hMsun)} & \multicolumn{3}{c}{$10^{12.0}-10^{15.0}$(\hMsun)}\\
\hline
 Simulation & $b/a$ & $c/a$ & $\theta$ & $b/a$ & $c/a$ & $\theta$\\
\hline
 MBII-25 & $0.779 \pm 0.004$ & $0.603 \pm 0.003$ & $31.151 \pm 0.603$ & $0.76 \pm 0.0195$ & $0.524 \pm 0.015$ & $31.785 \pm 3.525$ \\
 tsfr-High & $0.783 \pm 0.003$ & $0.618 \pm 0.003$ & $31.883 \pm 0.557$ & $0.777 \pm 0.02$ & $0.532 \pm 0.014$ & $32.512 \pm 3.490$ \\
 tsfr-Low & $0.775 \pm 0.004$ & $0.582 \pm 0.003$ & $33.306 \pm 0.648$ & $0.767 \pm 0.017$ & $0.533 \pm 0.013$ & $30.747 \pm 2.973$ \\
 AGN-High & $0.773 \pm 0.004$ & $0.604 \pm 0.003$ & $30.265 \pm 0.600 $ & $0.73 \pm 0.02$ & $0.513 \pm 0.013$ & $29.725 \pm 3.18$ \\
 Wind-High & $0.793 \pm 0.002$ & $0.636 \pm 0.002$ & $37.783 \pm 0.427$ & $0.747 \pm 0.02$ & $0.535 \pm 0.014$ & $32.512 \pm 3.49$ \\
\end{tabular}
\end{center}
\end{table*}

The cumulative shape distributions are plotted in Figure \ref{F:fig1_qs} in two mass bins of $10^{9.5-12.0}h^{-1}M_{\odot}$ and $10^{12.0-15.0}h^{-1}M_{\odot}$. Comparing the distributions and the mean values shown in Table~\ref{T:tab4}, we that in the lower mass bin the axis ratio $b/a$ is larger for the simulation with weaker wind feedback, although the effect is not large, within $2\sigma$ of the fiducial model - the deviation comparable to the difference between the mean of three MBII-25 runs and the original MBII-100 run.

Despite all deviations being moderate and not highly significant, some trends are nevertheless intriguing. For example, in the low mass bin weaker feedback makes galaxies rounder, while in the high mass bin the (mild) deviation is in the opposite direction. 

\begin{figure}
\begin{center}
\includegraphics[width=0.5\hsize]{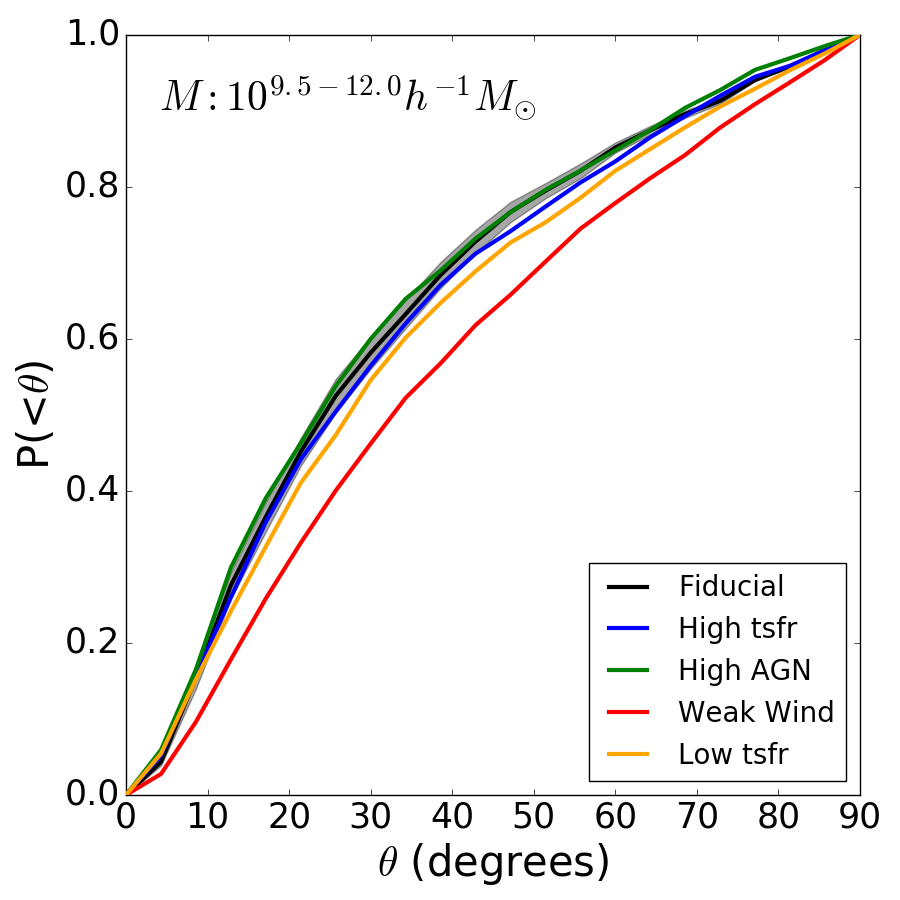}%
\includegraphics[width=0.5\hsize]{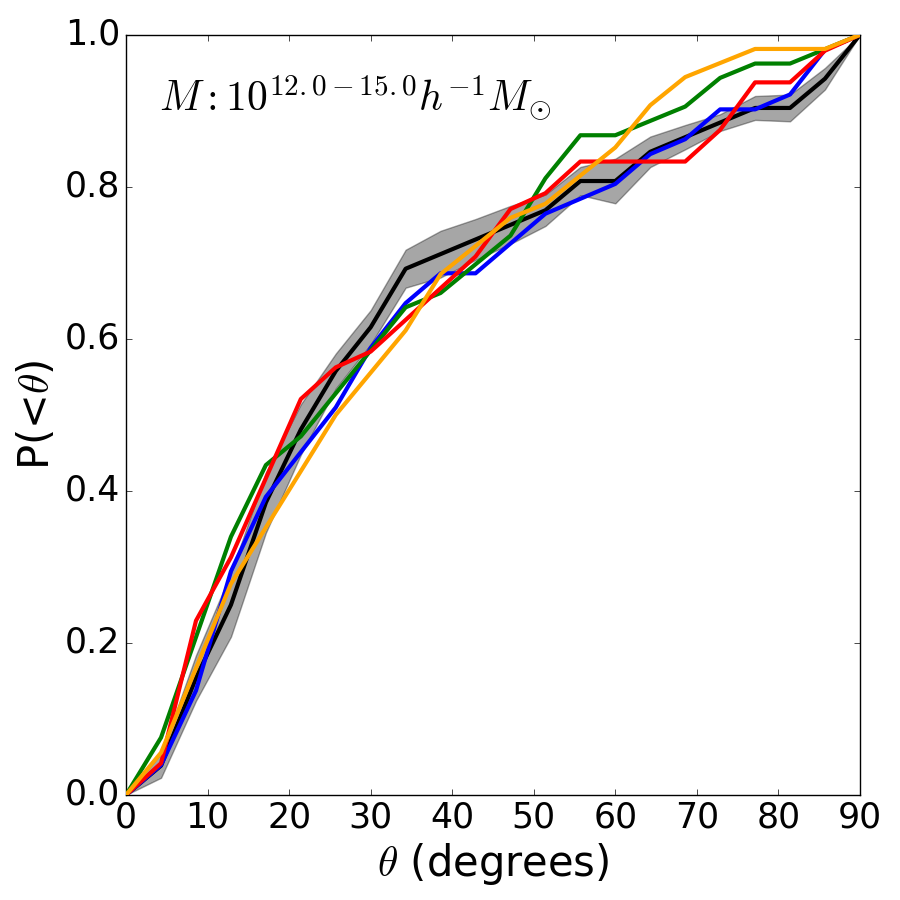}%
\caption{\label{F:fig_ma} Cumulative distribution function of the misalignment angles $\theta$ in two mass bins, $10^{9.5-12.0}h^{-1}M_{\odot}$  and $10^{12.0-15.0}h^{-1}M_{\odot}$,  of  several $25h^{-1}Mpc$ box simulations with varied physics. Black line with the gray band is the fiducial MBII-25 model and its error, shown with red lines in the previous section.
}
\end{center}
\end{figure}

The cumulative distributions of misalignment angles are shown in Figure \ref{F:fig_ma}. The effect of the lower wind loading factor is larger on the angles than on the shapes for lower mass galaxies - since the wind carries away linear and angular momenta, it can directly affect the orientation of the stellar distribution without affecting the shape that much. For more massive galaxies the effect disappears, however, as in that mass bin the feedback is dominated by AGN. One can hypothesize that AGN, being centrally located, are not able to eject large amounts of angular momentum. 

If such interpretation of our findings is valid, then the critical quantity that controls the distributions of shapes and angles is the angular momentum of the wind; once simulations get it right, their predictions for intrinsic alignment become robust and accurate.

In a previous study, \cite{2015MNRAS.453..721V} explored the variation in the ellipticities and misalignments, compared with their fiducial model of EAGLE simulation for three different feedback implementations. \cite{2015MNRAS.453..721V} investigated models with weaker and stronger stellar feedback and no AGN feedback. They also found that shapes are affected much less than angles, consistent with our hypothesis above. However, they find a larger effect of the stellar feedback on misalignment angles in more massive ($>10^{12}M_{\odot}$) galaxies, while their measurements for lower mass galaxies are too noisy to be conclusive. 

Overall, however, we find a good agreement with EAGLE simulations, which is encouraging, but not particularly surprising - modern simulations reproduce many observed properties of galaxies fairly.

\begin{figure}
\begin{center}
\includegraphics[width=0.5\hsize]{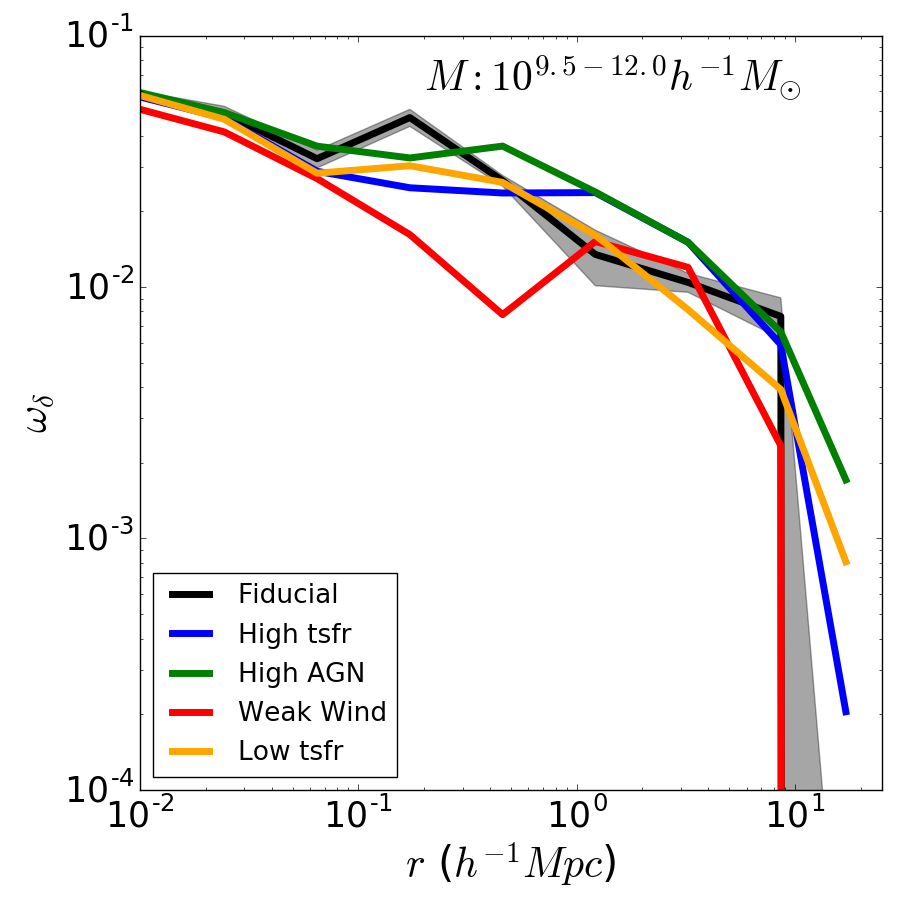}%
\includegraphics[width=0.5\hsize]{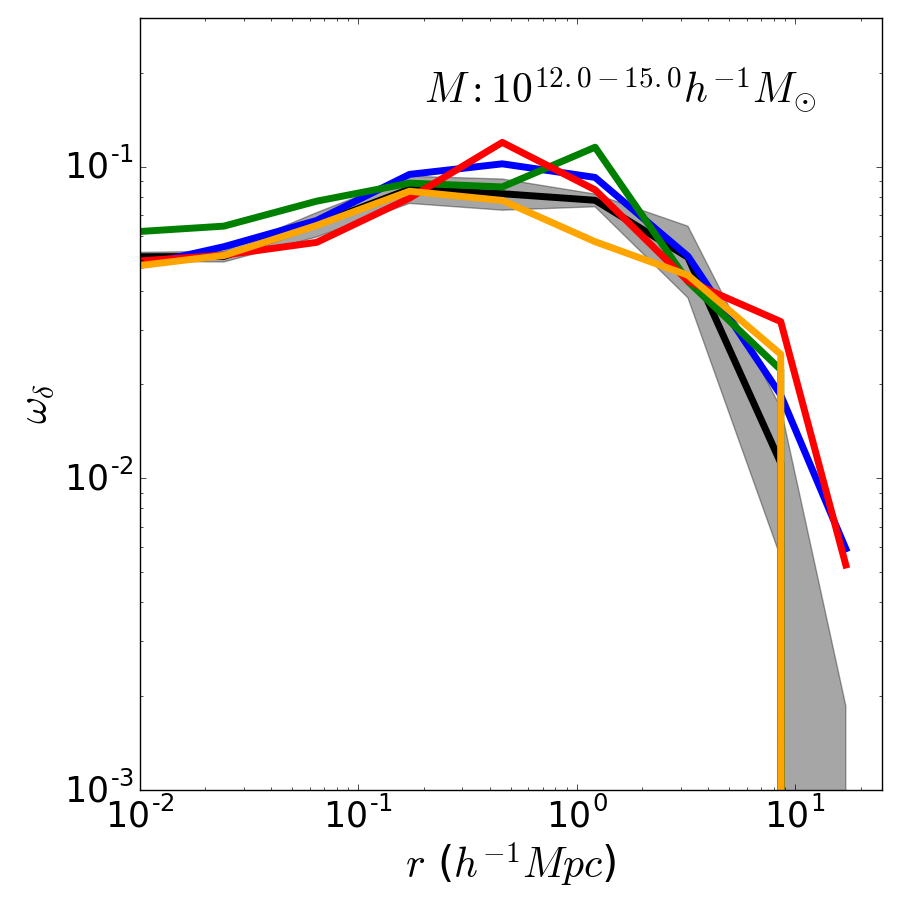}\newline%
\includegraphics[width=0.5\hsize]{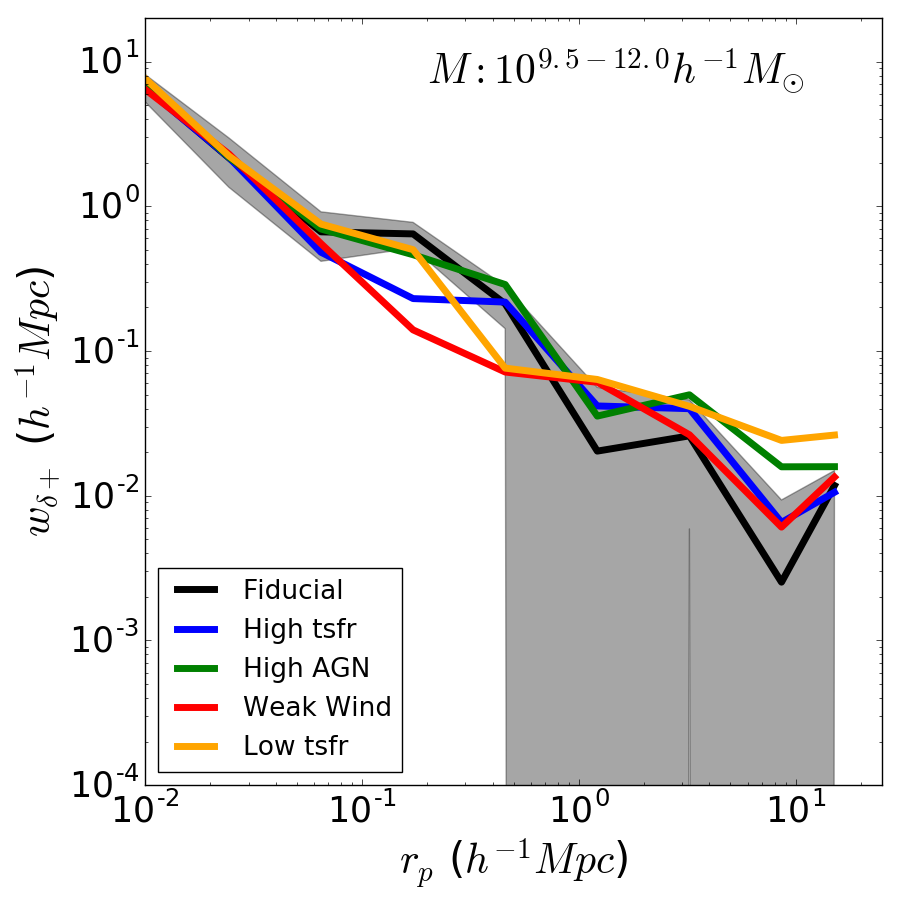}%
\includegraphics[width=0.5\hsize]{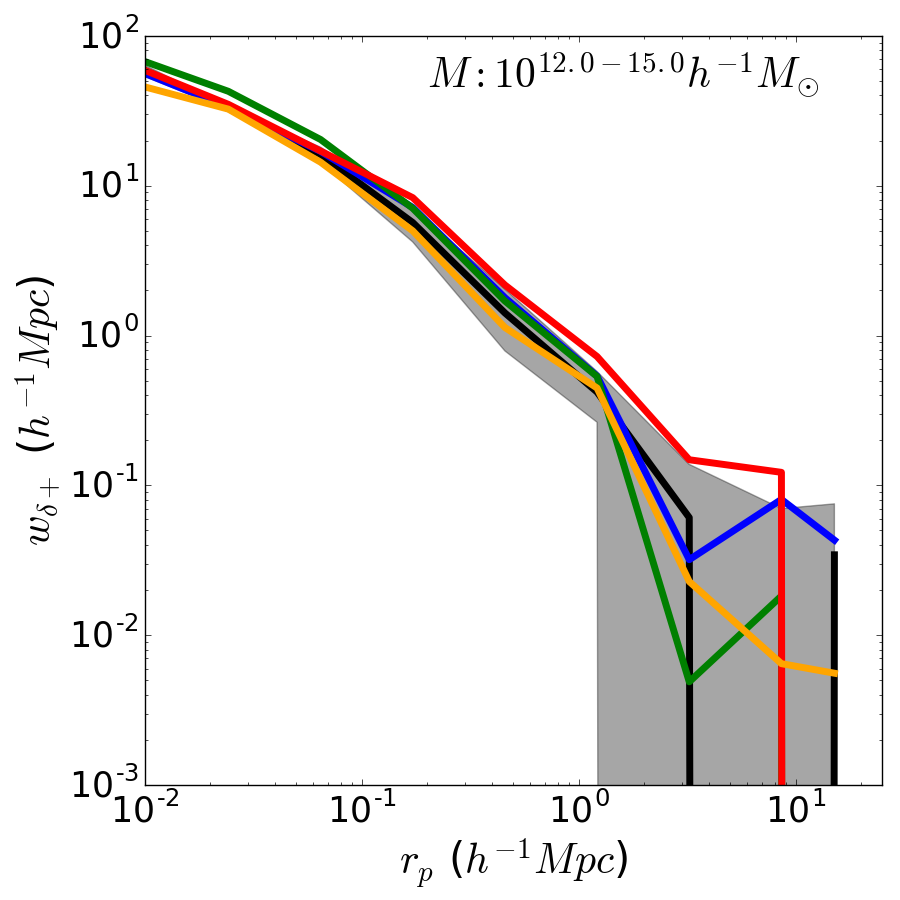}%
\caption{\label{F:fig4ed} ED and $w_{\delta +}$ correlation functions in two mass bins, $10^{9.5-12.0}h^{-1}M_{\odot}$  and $10^{12.0-15.0}h^{-1}M_{\odot}$, of several $25h^{-1}Mpc$ box simulations with varied physics. Black line with the gray band is the fiducial MBII-25 model and its error, shown with red lines in the previous section.
}
\end{center}
\end{figure}

Two point statistics for simulations with varied physics are shown in Figure \ref{F:fig4ed}. Differences between various models are similar to the level of difference between MBII-100 and MBII-25 runs: correlation functions agree well on small scales and in the high mass bin, but exhibit significant variations on scales between $0.1h^{-1}{\rm Mpc}$ and $1h^{-1}{\rm Mpc}$. These variations are non-monotonic and unsystematic, and are likely caused by the lack of statistics in our small box runs. However, just as in previous statistics, we find the largest difference in the run with the low wind mass loading factor.

In particular, the dip in the $w_\delta$ correlation function at $\sim 0.3h^{-1}{\rm Mpc}$ appears to be real - it is insensitive to the numerical details of computing the correlation function such as binning, sample selection, etc. The dip is located close to the radius where the one-halo term transitions to the two-halo term, and may reflect physical processes occurring at the halo-IGM interface. Unfortunately, our simulations volumes are too small to make any strongly statistically significant claim.

\section{Conclusions} \label{S:conclusions}

Our primary goal in this paper is to explore the effects of model parameters in the star formation and feedback models on the galaxy shapes and alignments using small volume simulations of size $25 h^{-1}Mpc$ on a side. As our fiducial model for the simulation, we adopted the same star formation and feedback model as in the MassiveBlack-II hydrodynamic simulation of galaxy formation \citep{2015MNRAS.450.1349K}, which is performed in a box of volume $(100h^{-1}Mpc)^3$.

Simulations with significantly (by factors of of 3 - 10) varying feedback show remarkable consistency with the fiducial run. Within the statistical precision we are able to achieve in our small volume runs, most of observational probes are insensitive to the details of subgrid physical modeling, with the exception of misalignment angles. We hypothesize that the angular momentum ejected by galactic winds is the most crucial physical quantity that determines the alignment of stellar shapes, and it remains one of the least robust quantities predicted in modern simulations of galaxy formation.

Our conclusions are also in good agreement with similar exploration of the role of subgrid physics on intrinsic alignments by the EAGLE simulation team.

\section*{Acknowledgments}

Fermilab is operated by Fermi Research Alliance, LLC, under Contract No.~DE-AC02-07CH11359 with the United States Department of Energy. AT is supported by the Fermilab Graduate Student Research Program in Theoretical Physics. Simulations have been performed on National Energy Research Supercomputing Center (NERSC) supercomputers ``Cori'' and ``Edison''. AT thanks Nishikanta Khandai and Tiziana DiMatteo also for sharing their P-Gadget code used to run the simulations in this work.
 
\bibliographystyle{apj}
\bibliography{psub5_v0}

\begin{thebibliography}{}
\expandafter\ifx\csname natexlab\endcsname\relax\def\natexlab#1{#1}\fi

\bibitem[{{Bernstein} \& {Jarvis}(2002)}]{2002AJ....123..583B}
{Bernstein}, G.~M., \& {Jarvis}, M. 2002, \aj, 123, 583

\bibitem[{{Blazek} {et~al.}(2015){Blazek}, {Vlah}, \&
  {Seljak}}]{2015JCAP...08..015B}
{Blazek}, J., {Vlah}, Z., \& {Seljak}, U. 2015, \jcap, 8, 015

\bibitem[{{Bridle} \& {King}(2007)}]{2007NJPh....9..444B}
{Bridle}, S., \& {King}, L. 2007, New Journal of Physics, 9, 444

\bibitem[{{Catelan} {et~al.}(2001){Catelan}, {Kamionkowski}, \&
  {Blandford}}]{2001MNRAS.320L...7C}
{Catelan}, P., {Kamionkowski}, M., \& {Blandford}, R.~D. 2001, \mnras, 320, L7

\bibitem[{{Chisari} {et~al.}(2015){Chisari}, {Codis}, {Laigle}, {Dubois},
  {Pichon}, {Devriendt}, {Slyz}, {Miller}, {Gavazzi}, \&
  {Benabed}}]{2015MNRAS.454.2736C}
{Chisari}, N., {Codis}, S., {Laigle}, C., {et~al.} 2015, \mnras, 454, 2736

\bibitem[{{Chisari} {et~al.}(2016){Chisari}, {Laigle}, {Codis}, {Dubois},
  {Devriendt}, {Miller}, {Benabed}, {Slyz}, {Gavazzi}, \&
  {Pichon}}]{2016arXiv160208373C}
{Chisari}, N.~E., {Laigle}, C., {Codis}, S., {et~al.} 2016, ArXiv e-prints,
  arXiv:1602.08373

\bibitem[{{Croft} \& {Metzler}(2000)}]{2000ApJ...545..561C}
{Croft}, R.~A.~C., \& {Metzler}, C.~A. 2000, \apj, 545, 561

\bibitem[{{Di Matteo} {et~al.}(2005){Di Matteo}, {Springel}, \&
  {Hernquist}}]{2005Natur.433..604D}
{Di Matteo}, T., {Springel}, V., \& {Hernquist}, L. 2005, \nat, 433, 604

\bibitem[{{Dubois} {et~al.}(2014){Dubois}, {Pichon}, {Welker}, {Le Borgne},
  {Devriendt}, {Laigle}, {Codis}, {Pogosyan}, {Arnouts}, {Benabed}, {Bertin},
  {Blaizot}, {Bouchet}, {Cardoso}, {Colombi}, {de Lapparent}, {Desjacques},
  {Gavazzi}, {Kassin}, {Kimm}, {McCracken}, {Milliard}, {Peirani}, {Prunet},
  {Rouberol}, {Silk}, {Slyz}, {Sousbie}, {Teyssier}, {Tresse}, {Treyer},
  {Vibert}, \& {Volonteri}}]{2014MNRAS.444.1453D}
{Dubois}, Y., {Pichon}, C., {Welker}, C., {et~al.} 2014, \mnras, 444, 1453

\bibitem[{{Genel} {et~al.}(2014){Genel}, {Vogelsberger}, {Springel}, {Sijacki},
  {Nelson}, {Snyder}, {Rodriguez-Gomez}, {Torrey}, \&
  {Hernquist}}]{2014MNRAS.445..175G}
{Genel}, S., {Vogelsberger}, M., {Springel}, V., {et~al.} 2014, \mnras, 445,
  175

\bibitem[{{Heavens} {et~al.}(2000){Heavens}, {Refregier}, \&
  {Heymans}}]{2000MNRAS.319..649H}
{Heavens}, A., {Refregier}, A., \& {Heymans}, C. 2000, \mnras, 319, 649

\bibitem[{{Heymans} {et~al.}(2006){Heymans}, {White}, {Heavens}, {Vale}, \&
  {van Waerbeke}}]{2006MNRAS.371..750H}
{Heymans}, C., {White}, M., {Heavens}, A., {Vale}, C., \& {van Waerbeke}, L.
  2006, \mnras, 371, 750

\bibitem[{{Hilbert} {et~al.}(2016){Hilbert}, {Xu}, {Schneider}, {Springel},
  {Vogelsberger}, \& {Hernquist}}]{2016arXiv160603216H}
{Hilbert}, S., {Xu}, D., {Schneider}, P., {et~al.} 2016, ArXiv e-prints,
  arXiv:1606.03216

\bibitem[{{Hirata} \& {Seljak}(2004)}]{2004PhRvD..70f3526H}
{Hirata}, C.~M., \& {Seljak}, U. 2004, \prd, 70, 063526

\bibitem[{{Jing}(2002)}]{2002MNRAS.335L..89J}
{Jing}, Y.~P. 2002, \mnras, 335, L89

\bibitem[{{Joachimi} {et~al.}(2013){Joachimi}, {Semboloni}, {Hilbert}, {Bett},
  {Hartlap}, {Hoekstra}, \& {Schneider}}]{2013MNRAS.436..819J}
{Joachimi}, B., {Semboloni}, E., {Hilbert}, S., {et~al.} 2013, \mnras, 436, 819

\bibitem[{{Katz} {et~al.}(1996){Katz}, {Weinberg}, \&
  {Hernquist}}]{1996ApJS..105...19K}
{Katz}, N., {Weinberg}, D.~H., \& {Hernquist}, L. 1996, \apjs, 105, 19

\bibitem[{{Khandai} {et~al.}(2015){Khandai}, {Di Matteo}, {Croft}, {Wilkins},
  {Feng}, {Tucker}, {DeGraf}, \& {Liu}}]{2015MNRAS.450.1349K}
{Khandai}, N., {Di Matteo}, T., {Croft}, R., {et~al.} 2015, \mnras, 450, 1349

\bibitem[{{Komatsu} {et~al.}(2011){Komatsu}, {Smith}, {Dunkley}, {Bennett},
  {Gold}, {Hinshaw}, {Jarosik}, {Larson}, {Nolta}, {Page}, {Spergel},
  {Halpern}, {Hill}, {Kogut}, {Limon}, {Meyer}, {Odegard}, {Tucker}, {Weiland},
  {Wollack}, \& {Wright}}]{2011ApJS..192...18K}
{Komatsu}, E., {Smith}, K.~M., {Dunkley}, J., {et~al.} 2011, \apjs, 192, 18

\bibitem[{{Krause} {et~al.}(2016){Krause}, {Eifler}, \&
  {Blazek}}]{2016MNRAS.456..207K}
{Krause}, E., {Eifler}, T., \& {Blazek}, J. 2016, \mnras, 456, 207

\bibitem[{{Laureijs} {et~al.}(2011){Laureijs}, {Amiaux}, {Arduini},
  {Augu{\`e}res}, {Brinchmann}, {Cole}, {Cropper}, {Dabin}, {Duvet}, {Ealet},
  \& et~al.}]{LAA+11}
{Laureijs}, R., {Amiaux}, J., {Arduini}, S., {et~al.} 2011, ArXiv e-prints,
  arXiv:1110.3193

\bibitem[{{Lee} {et~al.}(2008){Lee}, {Springel}, {Pen}, \&
  {Lemson}}]{2008MNRAS.389.1266L}
{Lee}, J., {Springel}, V., {Pen}, U.-L., \& {Lemson}, G. 2008, \mnras, 389,
  1266

\bibitem[{{LSST Science Collaboration} {et~al.}(2009){LSST Science
  Collaboration}, {Abell}, {Allison}, {Anderson}, {Andrew}, {Angel}, {Armus},
  {Arnett}, {Asztalos}, {Axelrod}, \& et~al.}]{LSST09}
{LSST Science Collaboration}, {Abell}, P.~A., {Allison}, J., {et~al.} 2009,
  ArXiv e-prints, arXiv:0912.0201

\bibitem[{{Mandelbaum} {et~al.}(2006){Mandelbaum}, {Hirata}, {Ishak}, {Seljak},
  \& {Brinkmann}}]{2006MNRAS.367..611M}
{Mandelbaum}, R., {Hirata}, C.~M., {Ishak}, M., {Seljak}, U., \& {Brinkmann},
  J. 2006, \mnras, 367, 611

\bibitem[{{Schaye} {et~al.}(2015){Schaye}, {Crain}, {Bower}, {Furlong},
  {Schaller}, {Theuns}, {Dalla Vecchia}, {Frenk}, {McCarthy}, {Helly},
  {Jenkins}, {Rosas-Guevara}, {White}, {Baes}, {Booth}, {Camps}, {Navarro},
  {Qu}, {Rahmati}, {Sawala}, {Thomas}, \& {Trayford}}]{2015MNRAS.446..521S}
{Schaye}, J., {Crain}, R.~A., {Bower}, R.~G., {et~al.} 2015, \mnras, 446, 521

\bibitem[{{Sirko}(2005)}]{2005ApJ...634..728S}
{Sirko}, E. 2005, \apj, 634, 728

\bibitem[{{Springel} {et~al.}(2005){Springel}, {Di Matteo}, \&
  {Hernquist}}]{2005MNRAS.361..776S}
{Springel}, V., {Di Matteo}, T., \& {Hernquist}, L. 2005, \mnras, 361, 776

\bibitem[{{Springel} \& {Hernquist}(2003)}]{2003MNRAS.339..289S}
{Springel}, V., \& {Hernquist}, L. 2003, \mnras, 339, 289

\bibitem[{{Tenneti} {et~al.}(2015{\natexlab{a}}){Tenneti}, {Mandelbaum}, \& {Di
  Matteo}}]{2015arXiv151007024T}
{Tenneti}, A., {Mandelbaum}, R., \& {Di Matteo}, T. 2015{\natexlab{a}}, ArXiv
  e-prints, arXiv:1510.07024

\bibitem[{{Tenneti} {et~al.}(2015{\natexlab{b}}){Tenneti}, {Singh},
  {Mandelbaum}, {Matteo}, {Feng}, \& {Khandai}}]{2015MNRAS.448.3522T}
{Tenneti}, A., {Singh}, S., {Mandelbaum}, R., {et~al.} 2015{\natexlab{b}},
  \mnras, 448, 3522

\bibitem[{{Velliscig} {et~al.}(2015){Velliscig}, {Cacciato}, {Schaye}, {Crain},
  {Bower}, {van Daalen}, {Dalla Vecchia}, {Frenk}, {Furlong}, {McCarthy},
  {Schaller}, \& {Theuns}}]{2015MNRAS.453..721V}
{Velliscig}, M., {Cacciato}, M., {Schaye}, J., {et~al.} 2015, \mnras, 453, 721

\bibitem[{{Vogelsberger} {et~al.}(2014{\natexlab{a}}){Vogelsberger}, {Genel},
  {Springel}, {Torrey}, {Sijacki}, {Xu}, {Snyder}, {Nelson}, \&
  {Hernquist}}]{2014MNRAS.444.1518V}
{Vogelsberger}, M., {Genel}, S., {Springel}, V., {et~al.} 2014{\natexlab{a}},
  \mnras, 444, 1518

\bibitem[{{Vogelsberger} {et~al.}(2014{\natexlab{b}}){Vogelsberger}, {Genel},
  {Springel}, {Torrey}, {Sijacki}, {Xu}, {Snyder}, {Bird}, {Nelson}, \&
  {Hernquist}}]{2014Natur.509..177V}
---. 2014{\natexlab{b}}, \nat, 509, 177

\end{thebibliography}

\end{document}